\documentclass[sigconf,nonacm]{acmart}

\usepackage{booktabs}
\usepackage{algorithm}
\usepackage{algpseudocode}
\usepackage{tikz}
\usetikzlibrary{arrows.meta,positioning,shapes.geometric}
\usepackage{graphicx}

\renewcommand\footnotetextcopyrightpermission[1]{}
\begin{document}

\title[Nearest but Not Dearest]{Nearest but Not Dearest:\\
Shared Curator-Feedback Infrastructure for Content-Only\\ Search and Recommendation}

\author{Matt Sandler}
\affiliation{%
  \institution{Feed.fm}
  \city{San Francisco}
  \state{California}
  \country{USA}
}
\email{matt@feed.fm}

\begin{abstract}
A deployed B2B music-discovery platform serves both query-driven search (semantic-text prompts, vibe-tag selection) and seed-driven recommendation (seed-track and artist stations) over one licensed catalog, one LAION-CLAP joint audio--text embedding space, one candidate-generation filter, and one ranking head --- and neither path consumes end-listener behavioral signal. In this content-only regime, professional curator judgment is the principal feedback signal available for maintaining the shared stack, and offline cosine similarity predicts it poorly: 38\% of cosine-nearest neighbors are rejected by curators. Inspecting the rejections reveals a clean partition: a majority (55\%) are \emph{sound} failures the encoder could in principle address (style, tempo, mood mismatch), and a substantial minority (37\%) are \emph{context} failures orthogonal to the waveform (wrong language, holiday content, devotional content, rights and lyric flags). We deploy this sound-vs-context decomposition as feedback infrastructure over the shared stack, routing each failure mode to the layer that can absorb it: context failures to a constraint filter at the candidate-generation layer, sound failures to an embedding reweighting head at the representation layer --- both layers sitting \emph{below} the search/recommendation split, so a single curator loop maintains both experiences. On 1{,}200 curator judgments collected over two production rounds one month apart, the combined intervention reduces rejection rate from 38.17\% to 28.83\% (\textbf{$-$24.5\% relative}, McNemar $\chi^2{=}22.4$, $p = 2.2 \times 10^{-6}$). An accounting decomposition attributes 4.08\,pp of the drop to filter-eligible categories and 5.25\,pp to non-filter categories; the deployment was unblinded and compound, so this is a production accounting bound, not a clean causal estimate. We present this as an industrial case study of a shared feedback loop rather than a validated general method, and close with lessons for unified discovery in the content-only regime: the failure partition is orthogonal to the paradigm partition, corrections land in layers shared by both paradigms, and a small in-house curator panel can maintain both when behavioral signal is unavailable.
\end{abstract}


\maketitle

\section{Introduction}\label{sec:intro}

A Spanish-language Latin-pop seed retrieves an English-language alt-pop track at cosine similarity $0.83$. The encoder captured the mid-tempo build, the bright vocal timbre, and the verse--chorus dynamics; a curator rejects it outright for \texttt{wrong\_language}. An English-language pop seed retrieves a Spanish-language ballad at $0.85$ in the opposite direction --- same airy production, same melodic register, rejected for the same reason. A modern electro-funk seed retrieves a traditional Christmas standard at $0.79$ --- same warm horns, same syncopated groove, rejected for \texttt{holiday}. The audio encoder did exactly what it was asked to do in each case; the nearest neighbor really is acoustically close. A curator rejects every one --- and in the content-only regime described below, curator judgment is the available proxy for listener suitability.

These rejections sit at a specific, named failure surface. A production music recommender embeds tracks as vectors and returns nearest neighbors --- a pipeline that is cheap, scales, and is increasingly common in production~\cite{schedl2018current}. Its candidates are, in practice, misaligned with curator suitability judgments more often than cosine similarity alone reveals. Nothing in this failure pattern is specific to our deployment: it is the predictable result of asking an audio encoder to satisfy judgments that partly live outside the audio, and any content-only system built on nearest-neighbor retrieval inherits it.

The system we report on is a deployed music-discovery stack, run by Feed.fm, a B2B licensed-music platform that powers third-party apps under a content-only regime: the retrieval stack learns from catalog content and curator judgment, not from listener interactions. Per-listener skip and like events do exist, but spread across many small third-party apps they are sparse and unevenly distributed, and the discovery stack described here does not consume them; there is no collaborative-filtering substrate beneath it. The stack is unified from the bottom up: a single licensed catalog, a single LAION-CLAP joint audio--text embedding space, a single candidate-generation filter, and a single ranking head serve both query-driven entry points (semantic-text prompts, vibe-tag selection) and seed-driven ones (seed-track and artist stations). The system serves four constituencies --- rightsholders licensing the catalog, third-party apps consuming discovery results, in-house curators producing judgment data, and listeners receiving the results --- with no behavioral signal from the last feeding the retrieval stack. This data regime is more common than the literature typically assumes --- DMCA-compliant radio, in-app ambient music, curated corporate audio, cold-start catalogs, and long-tail discovery all share it --- and it withholds from retrieval exactly the signal that most unified search-and-recommendation architectures lean on hardest: logged user interactions. In a content-only regime, curator judgment is not one signal among many; it is the principal judgment signal available, and it arrives as a single stream regardless of which entry point produced the result being judged. Recent industry evidence shows enriched metadata materially improves downstream listener behavior: a DDEX report on Amazon Music and Universal Music Group~\cite{ddex2022mead} measured a 4.6\% lift in streams (11.9\% on algorithmic stations) and a 7.8\% drop in skip rate from delivering richer genre and mood data on streaming platforms that \emph{do} have behavioral signal. The corresponding upstream question --- how a content-only system identifies and corrects nearest-neighbor failures before metadata is even attached --- has received less attention.

This paper reports a deployed answer. We inspected 458 thumbs-down judgments from 1{,}200 production recommendations and found a clean partition. A majority of rejections (55.2\%) are \emph{sound} failures the encoder could in principle address (style, tempo, mood mismatch). A substantial minority (36.9\%) are \emph{context} failures orthogonal to the waveform: wrong language, holiday content, devotional content, children's catalog, content-rights restrictions. The two groups share almost nothing: same high cosine similarity, same audio plausibility, totally different engineering levers.

\textbf{Contribution.} We propose and deploy a \emph{sound-vs-context failure decomposition} for embedding-based music retrieval, treated as feedback infrastructure rather than as an academic taxonomy. The decomposition routes each curator thumbs-down to a distinct production mechanism at a distinct layer of the shared stack: a constraint filter handles $\mathcal{C}_{\text{ctx}}$ at the candidate-generation layer, an embedding reweighting head handles $\mathcal{C}_{\text{sound}}$ at the representation layer. Both layers sit below the point where the stack forks into search-shaped and recommendation-shaped experiences, so corrections apply to every entry point; the curator evaluation in this paper measures the seed-track path. On the deployed system, the combined intervention reduces nearest-neighbor rejection rate from 38.17\% to 28.83\% --- a \textbf{$-$24.5\% relative drop}, McNemar paired $\chi^2 = 22.37$, $p = 2.2 \times 10^{-6}$ --- with an accounting decomposition attributing 4.08\,pp of the drop to filter-eligible ($\mathcal{C}_{\text{ctx}}$) categories and 5.25\,pp to non-filter ($\mathcal{C}_{\text{sound}} \cup \mathcal{C}_{\text{other}}$) categories. The cost asymmetry is the practical surprise: the constraint filter is days of catalog work, the reweighting head is weeks of model work, and the two mechanisms account for comparable shares of the gap.

\textbf{Scope of the claim.} This is an industrial case study, not a validated general method. The mechanisms are deliberately simple --- a SQL predicate and a linear projection --- and the evidence is one compound production deployment measured on the seed-track path by a four-curator panel. What we claim is narrower and, we think, more useful to practitioners: that the sound-vs-context partition is a sound routing criterion for expert feedback in a shared content-only stack, that the cheap half of the routing buys as much as the expensive half, and that the loop is operable by a small in-house team. \S\ref{sec:lessons} states what the design cannot separate and what measurements would.

The title's pun captures the failure pattern: cosine similarity finds neighbors that are nearest in audio space, but closeness in audio is not what makes a curator hold them dear.

\section{Related Work}\label{sec:related}

\textbf{Embedding-based music retrieval.} Audio encoders such as LAION-CLAP~\cite{wu2023laionclap}, MERT~\cite{li2024mert}, MULE~\cite{mccallum2022supervised}, and MusicFM~\cite{won2024foundation} are typically evaluated on tagging benchmarks~\cite{law2009mtat,bogdanov2019mtgjamendo}; direct nearest-neighbor usefulness against curator or listener judgment receives less attention.

\textbf{Beyond accuracy in recommender evaluation.} The RecSys community has long argued that accuracy metrics correlate imperfectly with what users actually want~\cite{mcnee2006accuracy,herlocker2004evaluating,cremonesi2010performance}. Music adds genre-, language-, and context-specific failure modes that have no clean analogue in news or e-commerce~\cite{schedl2018current,flexer2010effects,urbano2013eval}. Our contribution is concrete and operational: a partition of music-retrieval failures by which engineering lever resolves them.

\textbf{Unified search and recommendation.} Joint modeling of search and recommendation has been argued for on both theoretical and empirical grounds~\cite{zamani2018joint,zamani2020learning}, deployed industrially as a single contextual model serving both tasks~\cite{bhattacharya2024unicorn}, and renewed by generative retrieval over semantic IDs~\cite{rajput2023generative} --- where whether joint training helps or hurts each task remains an active question~\cite{penha2024bridging}. That line of work unifies primarily at the model layer and presumes logged user interactions from at least one paradigm. Our deployment unifies further down the stack --- shared catalog, shared joint audio--text embedding space, shared candidate filter --- and operates where \emph{neither} paradigm's retrieval draws on behavioral logs. The unification question then inverts: not how to co-train on joint logs, but how a single expert-feedback loop maintains retrieval quality for both entry points at once.

\textbf{Joint and multimodal search--recommendation models.} A second line of work unifies the two paradigms at the user-behavior layer, modeling interleaved search and browsing sequences in one model: USER~\cite{yao2021user} integrates the two behavior streams into a single sequence; SESRec~\cite{si2023sesrec} learns disentangled search representations to improve recommendation; UniSAR~\cite{shi2024unisar} models user transitions between the two modes. Multimodal sequential recommenders such as MISSRec~\cite{wang2023missrec} bring pretrained image--text encoders into the item tower so that content, not only IDs, carries the representation. Our system is closest in spirit to the latter --- a pretrained joint audio--text encoder is the item tower, and it is what lets text queries and seed tracks share one space --- but it has no user-behavior sequence to model, so the sequence-level unification of USER, SESRec, and UniSAR is unavailable. On the ranking side, cross-encoders~\cite{nogueira2019bert} and late-interaction models~\cite{khattab2020colbert} score a query--item pair jointly rather than through two independent vectors; we use only a bi-encoder with a learned projection, and return to cross-encoding as future work in \S\ref{sec:lessons}.

\textbf{Industry deployments and metadata-driven lift.} Recent RecSys 2025 work describes deployed audio recommenders driven by listener behavior at scale, including content-type calibration~\cite{spotifyCalibrated2025} and synthetic query generation for cold-start audiobooks~\cite{spotifyAudioBoost2025}. The DDEX/MEAD~\cite{ddex2022mead} report quantifies the downstream lift of richer genre and mood metadata on streaming platforms with behavioral signal. We complement these on the upstream side: how to identify and correct retrieval failures \emph{before} metadata enrichment, in a regime where retrieval has no listener-level signal to draw on.

\section{The Decomposition}\label{sec:method}

\subsection{Deployed system: one stack, four entry points}\label{sec:system}

The recommendable catalog comprises 57{,}758 rights-cleared tracks; a separate reference catalog of 98{,}789 popular tracks supports query-side seed lookup but is not served. Track audio is embedded with LAION-CLAP~\cite{wu2023laionclap} (HTSAT-base backbone, 512-d joint audio-text space, music-pretrained checkpoint). Vectors are stored in PostgreSQL with the \texttt{pgvector} extension; retrieval is exact cosine via a stored procedure. End-to-end query latency is 85\,ms mean and 201\,ms p95.

Four entry points share this stack in production, spanning the query-driven/seed-driven divide. Two are search-shaped: semantic-text prompt (decomposed by a lightweight LLM into keyword tokens, then CLAP-encoded into the joint audio--text space) and vibe-tag selection. Two are recommendation-shaped: seed-track and artist (mapped to the artist's mean embedding). All four resolve to the same operation --- nearest-neighbor retrieval in the shared CLAP space, over the same catalog, through the same candidate filter --- differing only in how the query vector is constructed. The joint audio--text encoder is what makes this collapse possible: a text prompt and a seed track land in the same space, so ``search'' and ``recommendation'' differ at the interface layer, not the retrieval layer. The curator evaluation in this paper uses seed-track queries exclusively; \S\ref{sec:lessons} discusses what the shared-stack structure implies for the other entry points.

A subset of context-aware rules pre-filters the candidate pool before cosine ranking; the rejections analyzed below survive that pre-filter. Contextual metadata used by both the pre-filter and the constraint filter we deploy in \S\ref{sec:intervention} --- holiday and seasonal flags, devotional tags, per-track language --- is sourced from third-party metadata and DDEX deliveries and stored at the audio-file level.

\subsection{The sound-vs-context partition}\label{sec:partition}

Curators use a structured rejection interface that records, for each thumbs-down, a single failure-mode code from the production schema. We partition the schema into two disjoint sets, plus an explicit residual:
\begin{align*}
\mathcal{C}_{\text{sound}} &= \{\texttt{tempo},\, \texttt{style},\, \texttt{mood}\}, \\
\mathcal{C}_{\text{ctx}}   &= \{\texttt{holiday},\, \texttt{wrong\_language},\, \texttt{explicit}, \\
                            &\quad\;\; \texttt{lyrics},\, \texttt{not\_popular},\, \texttt{bad\_quality}, \\
                            &\quad\;\; \texttt{childrens},\, \texttt{devotional}\}, \\
\mathcal{C}_{\text{other}} &= \{\texttt{other}\}.
\end{align*}
The partition is operational, not metaphysical: a code lands in $\mathcal{C}_{\text{ctx}}$ when the metadata that resolves it is non-acoustic and stored separately from the audio file (typically delivered by the rightsholder or surfaced from a tag dictionary). $\mathcal{C}_{\text{other}}$ captures codes the curator could not cleanly attribute; in the accounting analysis (\S\ref{sec:intervention}) we conservatively allocate $\mathcal{C}_{\text{other}}$ to the non-filter group alongside $\mathcal{C}_{\text{sound}}$. The three derived shares are:
\begin{equation}
\phi_{g} = \frac{\#\{i : c_i \in \mathcal{C}_{g}\}}{\#\{i : y_i = -\}} \quad \text{for } g \in \{\text{sound}, \text{ctx}, \text{other}\},
\end{equation}
with $\phi_{\text{sound}} + \phi_{\text{ctx}} + \phi_{\text{other}} = 1$.

\textbf{Why the partition matters.} The two halves of the partition are addressed by qualitatively different engineering work. $\mathcal{C}_{\text{sound}}$ rejections call for adjusting where the encoder places neighbors --- a representation-side intervention. $\mathcal{C}_{\text{ctx}}$ rejections do not respond to representation work at all: the audio is plausibly close, the rejection depends on metadata orthogonal to the waveform. They call for filtering, gating, or candidate-pool restriction --- catalog-infrastructure interventions. A retrieval system with a 38\% expert-rejection rate that does not know whether the rejections are sound or context will spend its budget on the wrong levers.

\subsection{Routing failures to mechanisms}\label{sec:routing}

\begin{figure*}[t]
  \centering
  \resizebox{\textwidth}{!}{%
  \begin{tikzpicture}[
    node distance=6mm and 12mm,
    box/.style={draw, rounded corners=2pt, minimum height=9mm, minimum width=24mm, align=center, font=\small},
    decision/.style={draw, diamond, aspect=2.4, minimum height=8mm, align=center, font=\small, inner sep=1pt},
    arrow/.style={-Latex, thick}
  ]
    \node[box] (seed) {Seed track};
    \node[box, right=of seed] (clap) {LAION-CLAP\\ \footnotesize embedding};
    \node[box, right=of clap] (filter) {Constraint\\ filter};
    \node[box, right=of filter] (rerank) {Reweighting\\ head};
    \node[box, right=of rerank] (recs) {Top-$k$\\ recs};
    \node[box, dashed, right=of recs, fill=gray!8] (curator) {Curator\\ judgment};
    \draw[arrow] (seed) -- (clap);
    \draw[arrow] (clap) -- (filter);
    \draw[arrow] (filter) -- (rerank);
    \draw[arrow] (rerank) -- (recs);
    \draw[arrow] (recs) -- (curator);

    \node[decision, below=10mm of rerank, fill=gray!8] (split) {$c \in \mathcal{C}_{\text{ctx}}?$};
    \draw[arrow] (curator.south) |- (split.east);
    \draw[arrow] (split.west) -| (filter.south)
      node[pos=0.25, above, font=\footnotesize] {yes};
    \draw[arrow] (split.north) -- (rerank.south)
      node[pos=0.5, right, font=\footnotesize] {no};
  \end{tikzpicture}%
  }
  \caption{The curator-feedback loop. A rejection categorized as $\mathcal{C}_{\text{ctx}}$ is routed to the constraint filter (catalog-side intervention); a rejection in $\mathcal{C}_{\text{sound}} \cup \mathcal{C}_{\text{other}}$ is routed to the reweighting head along the non-filter path (representation-side intervention). The two mechanisms address disjoint failure modes and update on different cadences.}\label{fig:loop}
  \Description{Block diagram of the production retrieval pipeline. A seed track flows into a LAION-CLAP embedding, then a constraint filter, then a reweighting head, then top-k recommendations, then a curator judgment. A decision diamond below routes context failures back to the constraint filter and sound or uncategorized failures back to the reweighting head.}
\end{figure*}
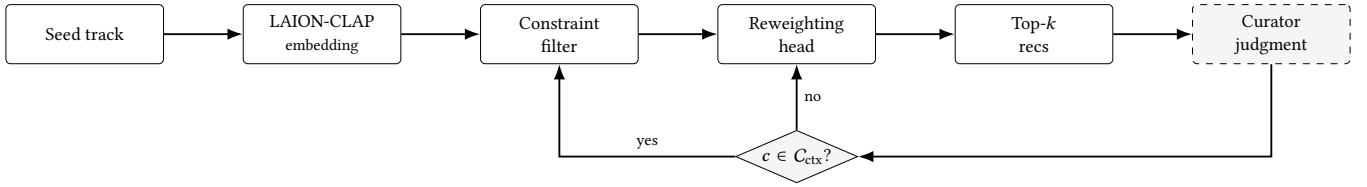

The deployed loop is shown in Figure~\ref{fig:loop}. Each curator thumbs-down with a code $c$ is dispatched as Algorithm~\ref{alg:routing}.

\begin{algorithm}[t]
\caption{Curator-feedback routing for a single thumbs-down}\label{alg:routing}
\begin{algorithmic}[1]
\Statex \textbf{Input:} seed $s$, retrieved neighbor $n$, failure-mode code $c$
\If{$c \in \mathcal{C}_{\text{ctx}}$}
  \State \textbf{constraint filter:} queue $(s, n, c)$ for catalog rule update
\Else \Comment{$c \in \mathcal{C}_{\text{sound}} \cup \mathcal{C}_{\text{other}}$ (non-filter path)}
  \For{each curator-accepted neighbor $n_+$ of seed $s$}
    \State \textbf{reweighting head:} emit triplet $(s, n_+, n)$
  \EndFor
\EndIf
\end{algorithmic}
\end{algorithm}

\textbf{Constraint filter (handles $\mathcal{C}_{\text{ctx}}$).} A rule-based filter applied at candidate generation, before cosine ranking. New $\mathcal{C}_{\text{ctx}}$ codes from curator feedback are batched into rule updates: a recurring \texttt{wrong\_language} flag against a specific source catalog drives a language-tag audit; a \texttt{holiday} flag during off-season prompts a seasonal-tag review; a \texttt{devotional} flag against a non-religious station prompts a devotional-tag rule. The filter is implemented as SQL predicates over the same metadata table that feeds the existing pre-filter; updates ship the same day they are reviewed. Cost is days of catalog work per rule cluster. Because the filter applies at candidate generation --- upstream of query-vector construction --- a rule learned from seed-track feedback (a corrected language tag, a seasonal flag) constrains text-prompt and vibe-tag retrieval as well.

\textbf{Reweighting head (handles $\mathcal{C}_{\text{sound}}$).} A standard metric-learning head trained on curator-derived positive/negative pairs. Curator thumbs-up pairs (seed, accepted neighbor) supply positives; thumbs-down pairs labeled $\mathcal{C}_{\text{sound}}$ supply negatives. The head is a single linear projection over the frozen 512-dimensional LAION-CLAP audio embedding, preserving dimensionality, $L_2$-normalized at output. Training uses a Euclidean triplet margin loss with margin $0.2$ over the $L_2$-normalized projections, optimized with AdamW (learning rate $10^{-4}$, weight decay $10^{-4}$, batch size 32 triplets) for up to 50 epochs with early stopping (patience 5) on a seed-level 80/20 train/validation split. The projection is initialized near identity and trained with an additional regularizer penalizing deviation from the identity; both choices counter overfitting, since a $512\times 512$ projection carries roughly $262$k parameters against the few hundred curator-derived triplets available. We construct triplets $(s, n_+, n_-)$ where $n_+$ is a curator-accepted neighbor of seed $s$ and $n_-$ is a $\mathcal{C}_{\text{sound}}$-rejected neighbor of the same seed, so the projection learns what curators consider a sonic match \emph{relative to that seed} rather than in absolute terms.

\textbf{Training-evaluation relationship.} The reweighting head was trained on the Round-1 curator feedback. The Round-2 aggregate is therefore a production deployment result on the same evaluation workload that fed the training feedback --- not a held-out generalization benchmark. To reduce leakage during model selection, we used a seed-level validation split: 20\% of training seeds were held out, with all their neighbors excluded from the training set. The paired Round-1$\to$Round-2 comparison reported in \S\ref{sec:intervention} therefore measures the change in curator rejection rate on the production evaluation workload \emph{after} feedback from that workload was incorporated into the reweighting head and the constraint filter. We report this as a production feedback-loop result rather than a held-out generalization benchmark; quantifying generalization to a never-judged seed population is left to future work.

At ranking time, retrieval uses cosine in the projected space only --- no blend with original cosine. The head is light enough to score candidates at query time without re-encoding the 57{,}758-track catalog; updates ship without recomputing the underlying CLAP embeddings. Cost is weeks of model and pipeline work per training cycle. The projection likewise lives in the shared space: any entry point whose query resolves to a CLAP vector can be scored through it, so a head trained on seed-track feedback is structurally available to the search-shaped paths; whether it improves them is a measurement question we return to in \S\ref{sec:lessons}.

The decomposition is what makes the routing well-defined. Both interventions are standard in isolation; the contribution here is the principled partition that decides which one each thumbs-down feeds.

\section{Evaluation}\label{sec:eval}

\subsection{Curator panel and protocol}

The panel comprises four professional in-house music curators with combined expertise spanning hip-hop, Latin music, pop, R\&B, classical, electronic dance, and metal. Each curator was assigned a disjoint slice of approximately 100 seed tracks (400 unique seeds total), sampled uniformly at random from the recommendable catalog. For each seed, the production recommender returned $k=3$ nearest neighbors, yielding $N = 1{,}200$ judgments per round. Round 1 was conducted on the baseline LAION-CLAP system (pre-filter only, no reweighting head); Round 2 was conducted approximately one month later, targeting the same 400-seed evaluation set after the constraint filter and reweighting head described in \S\ref{sec:routing} had been deployed. Curators recorded a binary thumbs-up / thumbs-down judgment and, for thumbs-downs, a single failure-mode code from the schema in \S\ref{sec:partition}.

\textbf{Paired-seed coverage.} The aggregate rejection-rate comparison in \S\ref{sec:intervention} uses the full $N = 1{,}200$ judgments per round. The seed-level paired test is restricted to the 269 of 400 seeds for which complete judgments exist in both rounds; the remaining 131 seeds had partial coverage (catalog availability, curator scheduling, or seeds that became ineligible under the new constraint filter and were therefore not served in Round 2). Reporting both statistics --- the marginal proportion comparison on all 1{,}200 judgments per round, and the paired McNemar test on the 269-seed paired subset --- guards against the residual concern that the marginal effect is driven by between-round seed-set differences.

\subsection{Baseline failure-mode breakdown}\label{sec:results-baseline}

Of $N = 1{,}200$ Round-1 recommendations, 458 (38.17\%) received a thumbs-down with a recorded failure-mode code; the remainder were accepted or did not have a code recorded. Table~\ref{tab:modes} reports the breakdown of the 458 rejections.

\begin{table}[t]
\centering
\small
\caption{Round-1 (baseline) rejection breakdown over $N = 1{,}200$ recommendations, $n = 458$ thumbs-downs with a recorded failure-mode code. Zero-count $\mathcal{C}_{\text{ctx}}$ categories (e.g., \texttt{bad\_quality}) are omitted from the table. \texttt{other} is reported as a separate row because the curator interface does not disaggregate sound-related from contextual ``other'' selections; the accounting analysis in \S\ref{sec:intervention} conservatively allocates $\mathcal{C}_{\text{other}}$ to the non-filter group alongside $\mathcal{C}_{\text{sound}}$.}\label{tab:modes}
\begin{tabular}{lrr}
\toprule
Failure mode & Count & \% rej. \\
\midrule
$\mathcal{C}_{\text{sound}}$ & \textbf{253} & \textbf{55.2\%} \\
\quad\texttt{style}             & 168 & 36.7\% \\
\quad\texttt{tempo}             & 52  & 11.4\% \\
\quad\texttt{mood}              & 33  & 7.2\% \\
\midrule
$\mathcal{C}_{\text{ctx}}$ & \textbf{169} & \textbf{36.9\%} \\
\quad\texttt{holiday}           & 68 & 14.8\% \\
\quad\texttt{wrong\_language}   & 67 & 14.6\% \\
\quad\texttt{explicit}          & 14 & 3.1\% \\
\quad\texttt{not\_popular}      & 8  & 1.7\% \\
\quad\texttt{childrens}         & 7  & 1.5\% \\
\quad\texttt{lyrics}            & 3  & 0.7\% \\
\quad\texttt{devotional}        & 2  & 0.4\% \\
\midrule
\texttt{other} (uncategorized)  & 36 & 7.9\% \\
\midrule
Total                           & \textbf{458} & \textbf{100\%} \\
\bottomrule
\end{tabular}
\end{table}

Both halves of the partition are substantial: $\phi_{\text{sound}} = 55.2\%$, $\phi_{\text{ctx}} = 36.9\%$, $\phi_{\text{other}} = 7.9\%$. Even under the conservative assumption that the entire \texttt{other} bucket is sound-related, $\phi_{\text{ctx}}$ remains a structurally significant share of failures and is unaddressable by encoder work alone.

\textbf{Worked examples.} High-similarity rejections in the Round-1 tail include: a synth-pop seed retrieving an obscure sound-alike at $\sigma = 0.927$ (\texttt{not\_popular}); a country-pop ballad retrieving a religious-country devotional at $\sigma = 0.892$ (\texttt{devotional}); a Spanish-language seed retrieving an English-language alt-pop track at $\sigma = 0.83$ (\texttt{wrong\_language}); a modern electro-funk seed retrieving a Christmas standard at $\sigma = 0.79$ (\texttt{holiday}). All four are sound-plausible; none are encoder-addressable.

\begin{figure}[t]
  \centering
  \includegraphics[width=\linewidth]{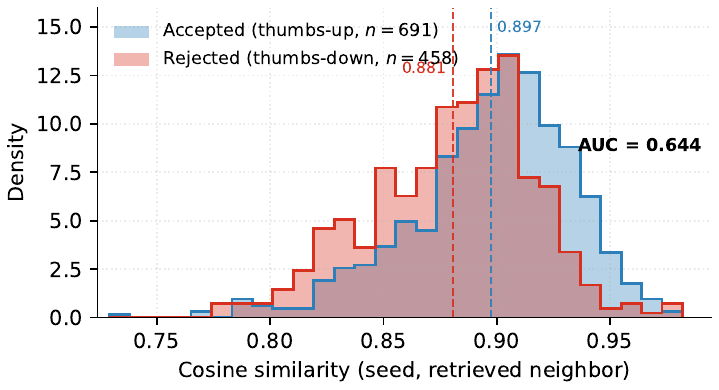}
  \caption{Cosine similarity distribution by curator judgment on Round-1 ($N=1{,}149$ judgments with a clear binary outcome and recorded failure-mode code). Accepted (thumbs-up, $n=691$) and rejected (thumbs-down, $n=458$) distributions overlap heavily; the small mean shift (0.897 vs.~0.881, dashed lines) yields AUC $=0.644$, the empirical basis for the partition-as-routing-criterion argument in \S\ref{sec:routing}.}\label{fig:cosineDensity}
  \Description{Overlaid histograms of cosine similarity between seed and retrieved neighbor, separated by curator judgment. The accepted (thumbs-up) distribution and the rejected (thumbs-down) distribution overlap substantially across the 0.80--0.95 range, with the accepted distribution shifted slightly higher.}
\end{figure}

\textbf{Cosine carries weak signal across categories.} As a sanity check on the partition's value, we treated cosine similarity as a binary classifier of curator judgment on Round-1, restricted to the $1{,}149$ judgments with both a binary outcome and (for rejections) a recorded failure-mode code (positive class \texttt{good\_match}: $n_+ = 691$ thumbs-up; negative class: $n_- = 458$ thumbs-down; the remaining 51 of the 1{,}200 judgments are excluded). The resulting AUC is $0.644$ (95\% bootstrap CI [0.613, 0.675]) --- a real lift over chance, but far too weak to act as a rejection detector on its own, and notably similar across rejection categories. Figure~\ref{fig:cosineDensity} shows the cosine similarity distribution by curator judgment; the two distributions overlap heavily despite a small mean shift ($\bar\sigma_{+} = 0.897$, $\bar\sigma_{-} = 0.881$). Sound-related rejection means ($\bar\sigma$: style 0.877, tempo 0.880, mood 0.880) fall in the same band as most context categories ($\bar\sigma$: holiday 0.883, \texttt{wrong\_language} 0.879, devotional 0.884). Cosine does not predict \emph{which} kind of rejection a curator will make; it only weakly predicts that one will occur. This is the empirical fact that makes the routing in \S\ref{sec:routing} necessary in the first place: the encoder cannot tell the two halves apart.

\subsection{Round-2 intervention}\label{sec:intervention}

Round 2 deployed the constraint filter and reweighting head described in \S\ref{sec:routing} on the same seed set. Table~\ref{tab:intervention} reports the headline result.

\begin{table}[t]
\centering
\small
\caption{Round-1 baseline vs.\ Round-2 (constraint filter + reweighting head) on $N = 1{,}200$ judgments per round. Two-proportion pooled $z = 4.84$, $p = 1.3 \times 10^{-6}$.}\label{tab:intervention}
\begin{tabular}{lrrrr}
\toprule
Metric & R1 & R2 & $\Delta$ & Rel. \\
\midrule
Seeds              & 400 & 400  & --- & --- \\
Judged recs        & 1{,}200 & 1{,}200 & --- & --- \\
Thumbs-downs       & 458 & 346  & $-112$ & --- \\
Rejection rate $\rho$ & 38.17\% & 28.83\% & $-9.33$\,pp & $\mathbf{-24.5\%}$ \\
\bottomrule
\end{tabular}
\end{table}

\textbf{Seed-level paired test.} Of the 269 seeds judged in both rounds, 76 improved (had thumbs-downs in R1, none in R2) versus 27 worsened (the inverse), a $2.81{:}1$ ratio. McNemar's $\chi^2$ on the paired discordant cells (continuity-corrected): $\chi^2 = 22.37$, $\mathrm{df}{=}1$, $p = 2.25 \times 10^{-6}$. The seed-level effect is consistent with the marginal proportion test in Table~\ref{tab:intervention} and is not driven by between-round seed-set differences. This treats the seed (not the individual $k=3$ recommendation) as the unit of analysis, addressing the within-seed clustering of the three retrieved candidates per query.

\textbf{Per-curator stability.} Each of the four curators judged a disjoint slice of approximately 100 seeds. Per-curator rejection rates moved consistently across rounds (all four curators within $\pm 5$\,pp of the aggregate change), ruling out a single-curator effect driving the headline drop.

\textbf{Accounting decomposition by category group.} Table~\ref{tab:perMech} reports the round-over-round breakdown by category group. The deployment was unblinded and the filter and reweighting head shipped together, so this is an arithmetic accounting across category groups, not a clean causal ablation. With that caveat, filter-eligible categories (\texttt{holiday}, \texttt{wrong\_language}, \texttt{explicit}, \texttt{not\_popular}, \texttt{childrens}, \texttt{lyrics}, \texttt{devotional}) drop from 169 thumbs-downs to 120 (49 fewer, 4.08\,pp of the aggregate). Non-filter categories ($\mathcal{C}_{\text{sound}} \cup \mathcal{C}_{\text{other}}$: \texttt{style}, \texttt{tempo}, \texttt{mood}, \texttt{other}) drop from 289 to 226 (63 fewer, 5.25\,pp of the aggregate). The two contributions sum to 9.33\,pp, matching the observed 9.33\,pp drop.

\begin{table}[t]
\centering
\small
\caption{Round-over-round accounting decomposition by category group. ``Filter-eligible'' contains $\mathcal{C}_{\text{ctx}}$; ``Non-filter'' contains $\mathcal{C}_{\text{sound}}$ and \texttt{other}. The deployment was unblinded and compound; this attribution is arithmetic across categories, not a controlled per-mechanism causal estimate.}\label{tab:perMech}
\begin{tabular}{lrrrr}
\toprule
Category group        & R1 TD & R2 TD & $\Delta$ & pp contrib. \\
\midrule
Filter-eligible ($\mathcal{C}_{\text{ctx}}$)         &  169 &  120 & $-49$  & $-4.08$\,pp \\
Non-filter ($\mathcal{C}_{\text{sound}} \cup \mathcal{C}_{\text{other}}$) &  289 &  226 & $-63$  & $-5.25$\,pp \\
\midrule
Total                                                  &  458 &  346 & $-112$ & $-9.33$\,pp \\
\bottomrule
\end{tabular}
\end{table}

\begin{figure}[t]
  \centering
  \includegraphics[width=\linewidth]{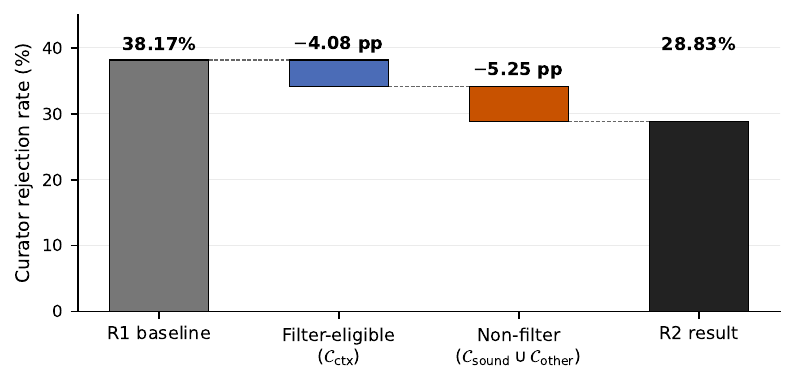}
  \caption{Rejection-rate waterfall, R1 baseline $\to$ R2 result. Filter-eligible categories ($\mathcal{C}_{\text{ctx}}$) account for $-4.08$\,pp of the drop; non-filter categories ($\mathcal{C}_{\text{sound}} \cup \mathcal{C}_{\text{other}}$) account for $-5.25$\,pp. The two mechanisms shipped together (unblinded compound deployment); this is an arithmetic accounting decomposition, not a controlled per-mechanism ablation.}\label{fig:permech}
  \Description{Waterfall chart of curator rejection rate. Round 1 baseline starts at 38.17 percent. A first step labeled "filter-eligible categories" lowers the rate by 4.08 percentage points to 34.09 percent. A second step labeled "non-filter categories" lowers the rate by a further 5.25 percentage points to 28.84 percent. The Round 2 result is shown as 28.83 percent.}
\end{figure}

\textbf{The cost asymmetry is the practitioner takeaway.} The constraint filter is days of catalog and rule work; the reweighting head is weeks of model and pipeline work. Under the accounting in Table~\ref{tab:perMech}, their contributions to the aggregate win are within 1\,pp of each other. A team that routes all curator-feedback work into a re-train cycle is leaving comparable gain on the table for the price of writing SQL predicates against a metadata table.

\textbf{What the accounting can and cannot separate.} Table~\ref{tab:perMech} splits the drop by \emph{rejection category}, not by \emph{mechanism}, and the two do not coincide. Leakage runs in both directions. The filter removes candidates before ranking, so the neighbors that surface in Round 2 are drawn from a different pool; a track that replaces a filtered-out holiday standard can itself be rejected for \texttt{style}, or accepted, which moves the non-filter count without the head having done anything. Conversely, the head reorders the survivors, and a re-ranked top-3 can surface or bury a \texttt{wrong\_language} track that the filter's rules did not yet cover, which moves the filter-eligible count without the filter having done anything. Two structural facts bound the leakage. First, the head is trained only on $\mathcal{C}_{\text{sound}}$ negatives, so it has no signal that would target context categories; any $\mathcal{C}_{\text{ctx}}$ movement it causes is incidental. Second, the filter is a deterministic predicate over metadata, so it cannot distinguish two candidates on sound; any $\mathcal{C}_{\text{sound}}$ movement it causes is through pool composition alone. The 4.08\,pp / 5.25\,pp split is therefore best read as ``the filter's share is at most the filter-eligible drop plus whatever pool-composition effect it had on the rest,'' and symmetrically for the head. Two measurements would close the gap and are the ones we would run first (\S\ref{sec:lessons}): a counterfactual replay of the Round-1 candidate lists through the Round-2 predicates, which the filter's determinism makes exact and cheap, and a staged rollout with a curator round between shipping the filter and shipping the head.

\subsection{Inter-rater agreement: a brief note}\label{sec:irr}

A two-curator pair re-judged $n=134$ shared seed-neighbor pairs in Round 1, yielding Cohen's $\kappa$~\cite{cohen1960coefficient} $= 0.451$ (95\% bootstrap CI $[0.30, 0.60]$; Landis-Koch~\cite{landis1977measurement} ``moderate''). Decomposing the 36 disagreements: 75\% are $\mathcal{C}_{\text{sound}}$ (\texttt{style} 50\%, \texttt{tempo} 19\%, \texttt{mood} 6\%), only 22\% are $\mathcal{C}_{\text{ctx}}$. Curators agree more readily on whether a track is contextually unsuitable than on whether a sonic match is good enough. Practically, $\phi_{\text{ctx}}$ is a more rater-robust signal than $\phi_{\text{sound}}$, and reweighting-head training data should account for the larger label-noise component on the sound side.

\section{Lessons and Limitations}\label{sec:lessons}

\textbf{The partition is the contribution; the mechanisms are commodity.} A constraint filter is a SQL predicate. A metric-learning head is a textbook component. What is not commodity is knowing which thumbs-down feeds which mechanism --- and that decision is exactly what the partition makes well-defined. Teams that bundle all feedback into a single retraining loop wash the two halves together and dilute both signals.

\textbf{$\phi_{\text{ctx}}$ is the cheaper half of the gain.} On our system, the filter-eligible group ($\mathcal{C}_{\text{ctx}}$) accounts for 4.08\,pp of the rejection-rate reduction at days-of-engineering cost. The non-filter group ($\mathcal{C}_{\text{sound}} \cup \mathcal{C}_{\text{other}}$, addressed by the reweighting head) accounts for 5.25\,pp at weeks-of-engineering cost. The cost-per-pp ratio favors the filter heavily. Practitioners with limited modeling capacity should route to $\mathcal{C}_{\text{ctx}}$ first.

\textbf{A small in-house panel is enough.} Four curators producing 1{,}200 judgments was sufficient to drive a deployable, statistically robust intervention without consuming listener behavioral data. The content-only regime is not a death sentence for measured improvement; it just shifts the judgment source from listeners to professionals. Curators are a proxy for listeners, and the proxy is unvalidated here: no behavioral loop is closed here to confirm that curator-approved retrieval is listener-preferred retrieval. We treat this as a structural property of the regime rather than a fixable gap, and note that the round-over-round \emph{drop} is robust to any fixed curator-listener offset.

\textbf{The failure partition is orthogonal to the paradigm partition.} Sound-vs-context is not search-vs-recommendation. A Christmas standard is wrong for a March seed track and equally wrong for a March text prompt; a tempo mismatch is a tempo mismatch regardless of how the query vector was constructed. Because both correction mechanisms live in layers shared by every entry point --- the filter at candidate generation, the head at representation --- feedback collected on one paradigm's workload structurally propagates to the other. This is the content-only analogue of cross-task transfer in unified models: transfer happens through shared infrastructure rather than through parameters co-trained on joint interaction logs. A team maintaining separate feedback loops per entry point would pay for the same correction up to four times; routing by failure mode instead of by entry point collects each correction once and amortizes it across the stack.

\textbf{Limitations.} (i) The Round-1$\to$Round-2 production deployment was unblinded and the intervention was compound (filter + head); the per-mechanism decomposition in \S\ref{sec:intervention} is an arithmetic bound, not a clean per-mechanism causal estimate. (ii) Scope is one deployed system on one catalog; generalization to other catalogs and other audio encoders is left to future work. (iii) Curators are in-house professionals; calibration drift relative to an external curator pool would limit generalization of the absolute rejection-rate levels, though the round-over-round drop is measured paired-seed and is robust to that drift. (iv) The four-curator panel is small; the inter-rater statistics make this size scrutable, but a larger pool is desirable. (v) Cross-entry-point transfer follows from the architecture (shared filter, shared representation) but is not yet measured: the curator protocol evaluated the seed-track path only, and the text-prompt and vibe-tag paths await an equivalent evaluation. (vi) The result is a change in curator rejection rate on the workload that fed the feedback, not a change on unseen seeds and not a change in listener satisfaction; the paper shows the feedback loop closes, not that it generalizes.

\textbf{Future work.} We list the measurements in the order we would run them.

\emph{Measure the search path.} The nearest measurement is the one limitation (v) names: rerun the \S\ref{sec:eval} protocol with query-driven entry points in place of seed tracks. The concrete protocol is already determined by the shared stack: for each of the 400 evaluation seeds, derive a text prompt (curator-written, or the seed's own vibe tags), retrieve $k=3$ through the same filter and head, and have the same panel judge with the same rejection schema. Because the filter is a predicate on metadata, its $\mathcal{C}_{\text{ctx}}$ corrections apply to text-prompt retrieval by construction; the open quantity is the head, whose projection was fit to seed-track triplets and may or may not improve neighbors of a text-derived query vector. That number --- how much of the seed-path lift transfers across the paradigm split for free --- is, to our knowledge, unmeasured in any content-only deployment, and it is the measurement that would turn the shared-infrastructure argument of this paper from structural into empirical.

\emph{Disentangle the mechanisms.} The counterfactual filter replay and the staged rollout described in \S\ref{sec:intervention} would replace the accounting bound with a per-mechanism estimate.

\emph{Add a third judge.} The panel is the bottleneck: 1{,}200 judgments per round is what four curators can produce alongside their regular work, and it is why Round 2 targets the same seeds rather than a held-out set. An LLM judge~\cite{zheng2023judging,thomas2024llmrel} is the obvious way to scale, but the partition predicts where it will and will not help. Context failures are resolved by metadata the model can read --- language, seasonal and devotional tags, explicit flags, lyrics --- and are the categories on which human curators already agree (\S\ref{sec:irr}); an LLM auditor over $\mathcal{C}_{\text{ctx}}$ could plausibly extend filter-rule discovery to the full catalog and to seeds no curator has judged. Sound failures are the categories on which curators disagree most, and a text-only judge does not hear the audio; an audio-capable judge would need to be calibrated against the panel on exactly the \texttt{style}/\texttt{tempo}/\texttt{mood} disagreements in \S\ref{sec:irr} before its labels could feed the head. We therefore see LLM judgment as a way to scale the cheap half of the loop first, and as a third rater whose agreement with the panel is itself a result worth reporting.

\emph{Cross-encode the sound side.} The reweighting head compresses each track into one vector and asks cosine to carry every aspect of the match. A cross-encoder~\cite{nogueira2019bert} over the (seed, candidate) pair --- or a late-interaction model~\cite{khattab2020colbert} over frame-level audio tokens --- can attend to the specific aspect on which two tracks agree or differ, which is what \texttt{tempo} and \texttt{mood} rejections at $\sigma > 0.88$ suggest a single vector cannot. The curator data is already in the right shape: every judgment is a labeled (seed, candidate) pair. The 85\,ms mean latency budget rules out cross-encoding the catalog, but not re-scoring a bi-encoder shortlist of a few dozen candidates, which is the standard retrieve-then-rerank arrangement. Whether a few hundred curator pairs suffice to train one is the open question; the identity-regularized linear head was chosen for the same data-scarcity reason.

\emph{Validate against listeners.} The discovery stack does not consume listener behavior, which is why curators stand in; skip and like events do exist, sparse and uneven across apps. Comparing aggregate skip rates on stations built from Round-2 versus Round-1 neighbors, where an app's volume permits, would test whether curator rejection tracks listener rejection, and is the only way to convert the result from an expert-proxy improvement into a listener-facing one.

\emph{Replicate across encoders.} Replicating the partition across additional encoders (including masked-token-trained alternatives such as MERT~\cite{li2024mert}) is a natural follow-up; whether the sound-vs-context split is paradigm-level or encoder-specific is an open empirical question.

\section{Conclusion}\label{sec:conclusion}

We described a sound-vs-context failure decomposition for embedding-based music retrieval and reported its deployment as curator-feedback infrastructure on a production discovery stack that serves search-shaped and recommendation-shaped entry points from one embedding space, one catalog, and one candidate filter. The partition routes each curator thumbs-down to one of two mechanisms --- a constraint filter for context failures or an embedding reweighting head for sound failures --- at two layers of the shared stack that sit below the paradigm split, on disjoint engineering cadences. The combined intervention reduces production rejection rate by 24.5\% relative ($p = 2.2 \times 10^{-6}$) on 1{,}200 judgments per round; an accounting decomposition over filter-eligible and non-filter category groups shows comparable reductions on each side, with a substantial cost asymmetry favoring the filter on a cost-per-percentage-point basis. The decomposition is the contribution; the mechanisms are off-the-shelf, and the evidence is a single production case study whose per-mechanism and cross-paradigm claims remain to be measured. For content-only discovery systems whose retrieval does not draw on behavioral signal in either paradigm, this is a low-overhead pattern for converting professional curator judgment into measurable retrieval improvement --- collected once, applied to every entry point the stack serves.

\section*{Ethics Statement}

The evaluation involved four in-house professional music curators compensated as part of regular employment; no external or crowd-sourced raters participated, and no end-listener data is used. The taxonomy names categories (e.g., \texttt{wrong\_language}, \texttt{devotional}, \texttt{childrens}) that, applied asymmetrically, could risk algorithmic bias against non-dominant languages, cultural traditions, or listener communities. The taxonomy is context-aware: a category fires only when a seed's context and a retrieved neighbor's properties mismatch, not as a standing judgment on any music category. We recommend practitioners audit protocol outputs by seed-context to detect and correct systematic under-service of non-dominant contexts.

\section*{AI Usage Statement}

AI assistants were used for grammar, style, and copy-editing of author-drafted prose, and for LaTeX formatting (table layout, figure placement, bibliography hygiene). All research ideas, experimental design choices, data analyses, statistical computations, and figures are the authors' own.

\bibliographystyle{ACM-Reference-Format}
\bibliography{references}

\end{document}